\documentclass[prd,superscriptaddress,showpacs,twocolumn,floatfix]{revtex4}

\usepackage{amsmath,amssymb}
\usepackage{graphicx}
\usepackage{bm}

\begin{document}

\title{Potential models and lattice correlators for quarkonia \\ 
  at finite temperature}
\author{W.M. Alberico}
\affiliation{Dipartimento di Fisica Teorica dell'Universit\`a di Torino and \\
  Istituto Nazionale di Fisica Nucleare, Sezione di Torino, \\ 
  via P.Giuria 1, I-10125 Torino, Italy}
\author{A. Beraudo}
\affiliation{ECT*, strada delle Tabarelle 286, I-38050 Villazzano (Trento),
  Italy}
\author{A. De Pace}
\affiliation{Dipartimento di Fisica Teorica dell'Universit\`a di Torino and \\
  Istituto Nazionale di Fisica Nucleare, Sezione di Torino, \\ 
  via P.Giuria 1, I-10125 Torino, Italy}
\author{A. Molinari} 
\affiliation{Dipartimento di Fisica Teorica dell'Universit\`a di Torino and \\
  Istituto Nazionale di Fisica Nucleare, Sezione di Torino, \\ 
  via P.Giuria 1, I-10125 Torino, Italy}

\begin{abstract}
We update our recent calculation of quarkonium Euclidean correlators at finite
temperatures in a potential model by including the effect of zero modes in the
lattice spectral functions. These contributions cure most of the previously
observed discrepancies with lattice calculations, supporting the use of
potential models at finite temperature as an important tool to complement
lattice studies.
\end{abstract}



\maketitle

Quarkonia at finite temperature are an important tool for the study of
quark-gluon plasma formation in heavy ion collisions (see, e.~g.,
Ref.~\cite{Sat07}). 
Many efforts have been devoted to determine the dissociation temperatures of
$Q\bar{Q}$ states in the deconfined medium, using either lattice calculations
of quarkonium spectral functions
\cite{Asa03,Asa04,Dat04,Ume05,Iid06,Jak07,Aar07} or non-relativistic
calculations based upon some effective (screened) potential
\cite{Dig01,Shu04,Won05,Alb05,Won07,Cab06,Alb07,Moc07}.

Lattice studies are directly based on quantum chromodynamics and should
provide, in principle, a definite answer to the problem.
However, in lattice studies the spectral functions have to be extracted ---
using rather limited sets of data --- from the Euclidean (imaginary time)
correlators, which are directly measured on the lattice.
This fact, together with the intrinsic technical difficulties of lattice
calculations, somehow limits the reliability of the results obtained so far,
and also their scope, which in fact is essentially limited to the 
mass of the ground state in each $Q\bar{Q}$ channel.

Potential models, on the other hand, provide a simple and intuitive framework
for the study of quarkonium properties at finite temperature, allowing one to
calculate quantities that are beyond the present possibilities for lattice
studies. The main problem  of the latter approach is the determination of the
effective potential: although at zero temperature the use of effective
potentials and their connection to the underlying field theory is well
established, at finite $T$ the issue is still open \cite{Sim05,Lai07a,Lai07b}.

Calculations of the $c\bar{c}$ and $b\bar{b}$ dissociation temperatures, using
different potential models based upon the lattice free and internal energies,
have found on the whole a reasonable agreement with the results from the
lattice studies \cite{Won05,Alb05,Won07,Alb07}. On the other hand,
calculations of Euclidean correlators using a variety of potential models were
not able to reproduce the temperature dependence of the lattice correlators 
\cite{Moc06,Cab06,Alb07}.

In Ref.~\cite{Alb07} it was pointed out that a precise quantitative
agreement with the lattice correlators should not be expected, mainly because
lattice spectral functions are strongly affected in the continuum region by
artifacts due to the finite lattice spacing. 
Actually, while some degree of qualitative agreement had been found for the
$S$-wave correlators, it was somehow disturbing the finding that for the
$P$-wave correlators the temperature dependence of the potential model was even
qualitatively different from the lattice one.
In Ref.~\cite{Alb07} this different behavior has been shown to be due to a
sizable amount of strength at low energy that was present in the $P$-wave
lattice spectral functions, but not in the potential model estimate.
This sizable amount of spectral strength --- which is present below the
$Q\bar{Q}$ threshold even when no bound states are supported --- strictly
speaking goes beyond a potential model description. Hence, at very low energy
the effective potential approach has to be supplemented by different
mechanisms. 

Recently, in Ref.~\cite{Ume07} it has been shown that the lattice calculations
of meson correlators at finite temperature contain a constant contribution, due
to the presence of zero modes in the spectral functions.
While the presence of a zero mode in the vector channel had already been
discussed in the literature, in the $P$-wave channels it had generally been
overlooked. 

In the following, we want to show that the inclusion of the zero mode
contributions in the calculations of Ref.~\cite{Alb07} can reconcile potential
model and lattice correlators. 
These contributions have also been considered in Ref.~\cite{Moc07} and we shall
comment below on the results obtained there. An alternative explanation of the
potential model vs lattice discrepancy has been proposed in Ref.~\cite{Won07}
and, again, we defer at the end our comments.

Here we briefly outline the model developed in Refs.~\cite{Alb05,Alb07}, to
which we refer the reader for all the details.
The object we are interested in is the $Q\bar{Q}$ Euclidean time correlator at
finite temperature $T$, defined as the thermal expectation value of a hadronic
current-current correlation function in Euclidean time $\tau$ for a given
mesonic channel $H$ (see, e.~g., Ref.~\cite{Bra05}, Chap.~7): 
\begin{equation}
\label{eq:GH}
  G_H(\tau,T) = \langle j_H(\tau) j_H^\dagger(0) \rangle,
\end{equation}
where $j_H=\bar{q}\Gamma_H q$ and $\Gamma_H=1$, $\gamma_5$, $\gamma_\mu$,
$\gamma_\mu\gamma_5$. The four vertex operators $\Gamma_H$ correspond,
respectively, to the scalar, pseudoscalar, vector and axial-vector mesonic
channels, which in turn, at zero temperature, correspond to the $\chi_{c0}$
($\chi_{b0}$), $\eta_c$ ($\eta_b$), $J/\Psi$ ($\Upsilon$) and $\chi_{c1}$
($\chi_{b1}$) quarkonium states for the $c\bar{c}$ ($b\bar{b}$) system,
respectively. In some lattice studies only the spatial components in the vector
and axial-vector channels are considered
($\Gamma_H=\gamma_i,\gamma_i\gamma_5$).
Moreover, we shall restrict ourselves, as in most lattice calculations, to the
case of spatial momentum $p=0$.

The correlators of Eq.~(\ref{eq:GH}) are related to the corresponding spectral
functions through an integral transform,
\begin{equation}
\label{eq:GHs}
  G_H(\tau,T) = \int_0^\infty d\omega\,\sigma_H(\omega,T) K(\tau,\omega,T),
\end{equation}
which is regulated by the temperature kernel
\begin{equation}
  K(\tau,\omega,T) = \frac{\cosh[\omega(\tau-1/2T)]}{\sinh[\omega/2T]}.
\end{equation}
The temperature dependence of the correlators is usually studied by introducing
the ratio between the correlation function $G_H(\tau,T)$ at some temperature
$T$ and the so-called {\em reconstructed correlator},
\begin{equation}
  G_H^{\text{rec}}(\tau,T,T_r) = \int_0^\infty d\omega\,\sigma_H(\omega,T_r) 
    K(\tau,\omega,T),
\end{equation}
calculated using the kernel at the temperature $T$ and the spectral function at
some reference temperature $T_r$. This procedure should eliminate the trivial
temperature dependence due to the kernel and differences from one in the ratio
should then be ascribed to the temperature dependence of the spectral function.

In the potential model of Ref.~\cite{Alb07}, the spectral function has been
expressed as 
\begin{equation}
\label{eq:sigmaHPM}
  \sigma_H(\omega,T) = \sum_n F_{H,n}^2 \delta(\omega-M_n) + \theta(\omega-s_0) 
    F_{H,\omega-s_0}^2, 
\end{equation}
where in the right hand side the sum over $n$ runs over the bound states of
mass $M_n$ and the last term represents the continuum contribution, $s_0$ being
the continuum threshold; $F_{H,n}^2$ and $F_{H,\epsilon}^2$ are the couplings
associated to the discrete and continuum states, respectively, and they can be
expressed \cite{Bod95,Moc06} in terms of the wave function at the origin for
the $S$-states (pseudoscalar and vector channels),
\begin{subequations}
\begin{equation}
  F^2_{PS} = \frac{N_c}{2\pi} |R(0)|^2 \quad \textrm{and} \quad
    F^2_{V} = \frac{3N_c}{2\pi} |R(0)|^2,
\end{equation}
and in terms of the first derivative of the wave function at the origin for the
$P$-states (scalar and axial-vector channels),
\begin{equation}
  F^2_{S} = \frac{9N_c}{2\pi m^2} |R'(0)|^2 \quad \textrm{and} \quad
    F^2_{A} = \frac{9N_c}{\pi m^2} |R'(0)|^2,
\end{equation}
\end{subequations}
$N_c$ being the number of colors and $m$ the quark mass.
Since the integration over the energy in Eq.~(\ref{eq:GHs}) can reach very high
excitation energies, the wave functions have been renormalized to account for
relativistic kinematical effects, that is the spectral functions have the
correct asymptotic dependence ($\sim\omega^2$).

The potential that we employ to generate the $Q\bar{Q}$ wave functions
\cite{Alb05,Alb07} is given by a linear combination \cite{Won05} of the
color-singlet free and internal energies obtained on the lattice from the
Polyakov loop correlation functions \cite{Kac02,Kac04,Kac05,Pet04}.
The only partially free parameter one has in this approach is the bare heavy
quark mass \footnote{Although one in principle would like to use the physical
quark masses, it should be remembered that the lattice free energies are
calculated for static (infinite mass) heavy quarks.}: for instance, using the
physical mass for the quark $c$, the $J/\psi$ dissociation temperature occurs
below $1.5T_c$, whereas with slightly heavier quarks it can be brought above
$1.5T_c$ \cite{Alb07}.

When addressing the Euclidean correlators, one should also account for the
presence of zero mode contributions to the spectral functions at finite
temperature. There is no zero mode in the pseudoscalar channel, whereas the
vector channel has been discussed in Refs.~\cite{Moc06,Pet06}.
Recently, is has been pointed out that these modes provide an important
contribution also in the lattice scalar and axial-vector correlators
\cite{Ume07}. 
In the non-interacting case, they yield a term proportional to $\delta(\omega)$
in the spectral function and a $\tau$-independent term in the 
Euclidean 
correlator: 
\begin{subequations}
\label{eq:sigmaG0}
\begin{eqnarray}
  \sigma_H^{(0)} &=& \chi_H(T)\omega\delta(\omega) \\
\label{eq:G0}
  G_H^{(0)}      &=& T \chi_H(T).
\end{eqnarray}
\end{subequations}
In the interacting theory, one may assume the delta function to be smeared to a
Lorentzian, with a width related to the heavy quark diffusion constant
\cite{Pet06}.

Here, for simplicity, we employ the free expressions of Eq.~(\ref{eq:sigmaG0}),
where all the susceptibilities $\chi_H$ can be expressed in terms of a scalar
piece, 
\begin{subequations}
\label{eq:suscS0}
\begin{equation}
  \chi_S(T) = -4 N_c \int\frac{d\bm{k}}{(2\pi)^3} \frac{m^2}{\omega_k^2}
    \frac{\partial n_F}{\partial\omega_k},
\end{equation}
and a charge piece,
\begin{equation}
  \chi_0(T) = -4 N_c \int\frac{d\bm{k}}{(2\pi)^3} 
    \frac{\partial n_F}{\partial\omega_k},
\end{equation}
\end{subequations}
as 
\begin{eqnarray}
  \mathrm{pseudoscalar:}&& \chi_{PS}=0 \nonumber \\
  \mathrm{scalar:}      && \chi_S      \nonumber \\
  \mathrm{vector(}\gamma_\mu\mathrm{):} && \chi_V=-\chi_S \nonumber \\
  \mathrm{vector(}\gamma_i\mathrm{):}   && \chi_V=\chi_0-\chi_S \nonumber \\
  \mathrm{axial(}\gamma_\mu\gamma_5\mathrm{):} && \chi_A=3\chi_S \nonumber \\
  \mathrm{axial(}\gamma_i\gamma_5\mathrm{):}   && \chi_A=\chi_0+2\chi_S. 
\end{eqnarray}
In Eq.~(\ref{eq:suscS0}), $\omega_k=\sqrt{k^2+m^2}$ and
$n_F=1/[\exp(\omega_k/T)+1]$. 

The zero mode contributions to the spectral function arise from
processes in which the meson current, rather than inducing transitions
resulting in the creation-annihilation of a $Q\bar{Q}$ pair, is absorbed by a
heavy (anti-)quark of the thermal bath, which is in turn scattered with a
sligthly modified momentum. The detailed balance between this process and its
inverse results in the factors $\partial n_F/\partial\omega_k$ in
Eq.~(\ref{eq:suscS0}). 

\begin{figure}
\includegraphics[clip,width=0.45\textwidth]{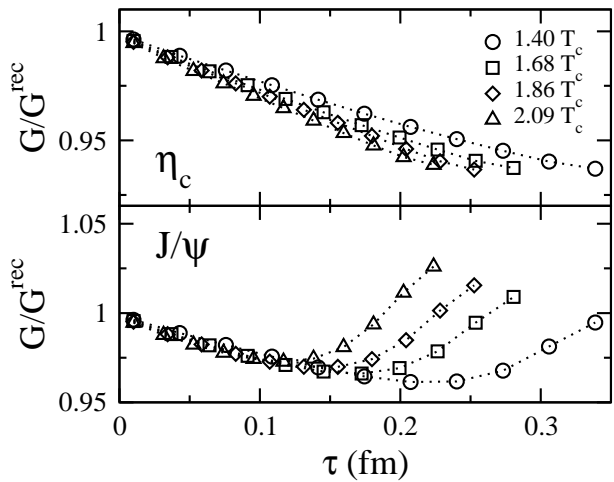}
\caption{ Ratio of the charmonium $S$-wave Euclidean correlators at various
  temperatures to the reconstructed one at $T=1.05T_c$: pseudoscalar
  (upper panel) and vector (lower panel) channels.
}
\label{fig:GratioSNf2}
\end{figure}

\begin{figure}
\includegraphics[clip,width=0.45\textwidth]{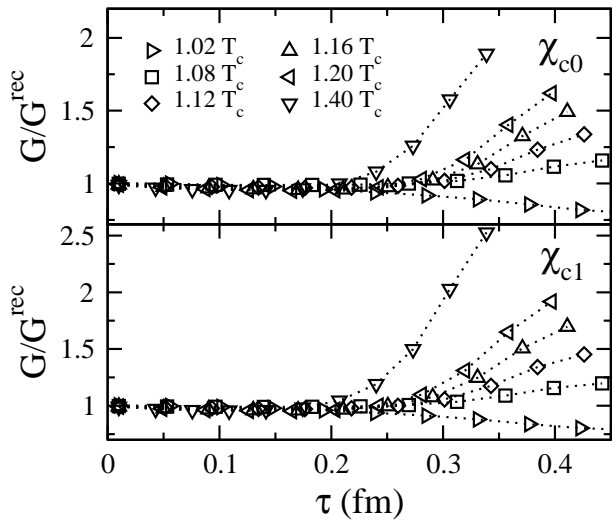}
\caption{ As in Fig.~\protect\ref{fig:GratioSNf2}, but for the $P$-wave
  correlators: scalar (upper panel) and axial-vector (lower panel) channels.
}
\label{fig:GratioPNf2}
\end{figure}

In Figs.~\ref{fig:GratioSNf2} and \ref{fig:GratioPNf2} we display our results
for $G_H(\tau,T)/G_H^{\text{rec}}(\tau,T,T_r)$ for charmonium at $T_r=1.05 T_c$
and $N_f=2$. 
They should be compared to the lattice calculations of Ref.~\cite{Aar07} (their
Figs.~3 and 4, respectively). They should also be compared to the results of
Ref.~\cite{Alb07}, where the present model has been employed without including
the zero mode contributions (note that in a pure potential model calculation
the ratio $G/G^{\text{rec}}$ depends only on the angular momentum state, $S$ or
$P$-wave, and not on the specific channel: thus, for instance, in
Ref.~\cite{Alb07} the ratio in the vector channel coincides with the
pseudoscalar one). 

The temperature dependence of the ratio is now well reproduced in the scalar
and axial-vector channels, thanks to the constant contribution of
Eq.~(\ref{eq:G0}). In the vector channel the strong enhancement of the ratio at
large $\tau$ is also well described, owing again to the zero mode term in the
spectral function. On the whole, the data of Ref.~\cite{Aar07} are described
semi-quantitatively: we stress again, as thoroughly discussed in
Ref.~\cite{Alb07}, that an accurate description of the lattice data should not
be expected, due to the discretization effects in the lattice calculations.

In the pseudoscalar channel, where no constant contribution to the correlators
is present, the ratio evaluated on the lattice tends (also in the quenched
case) to stay around one up to some temperature (interpreted as the
dissociation one) and then it starts decreasing; in our calculation, on the
other hand, the ratio is monotonically decreasing with $\tau$ at all the
temperatures. A remnant of this behavior is also visible in the vector channel,
where the ratio drops slightly below one, before jumping up.

Indeed, the binding due to the potential extracted from the lattice free
energies gets softer with increasing $T$ and this is reflected into the
decrease of the square wave function at the origin \cite{Alb07}.
On the other hand, the lattice spectral functions generally display ground
state peaks of nearly constant strength up to the dissociation temperature.
Note that when in the potential model the bound state energy is going to zero,
a strong resonance-like \footnote{Strictly speaking there are no $S$-wave
resonances for potentials without a repulsive barrier; however, the large
enhancements in the cross-section associated to nearly bound states are often
called zero-energy resonances \protect\cite{Joa75}.} contribution appears in
the continuum (see the spectral functions displayed in Ref.~\cite{Alb07}).
Although it might be possible that on the lattice this contribution has been
mistakenly identified as a bound state, in our calculations it still has a
strength gradually decreasing with $T$.

Note, anyway, that the trend predicted in the potential model is qualitatively
correct, since the quenching of the pseudoscalar ratio gets stronger with
increasing $T$. 
Similar results are also obtained for the quenched calculations of
Refs.~\cite{Dat04,Jak07} and we do not display them here.

In Ref.~\cite{Won07} an alternative explanation for the flatness, up to the
dissociation temperature, of the $G/G^{\text{rec}}$ ratio in the pseudoscalar
channel has been proposed: in that potential model no continuum contribution is
present when there are $S$-wave bound states; thus, assuming the existence of
just one bound state, one should have
$G_{PS}(\tau,T)/G_{PS}^{\text{rec}}(\tau,T,T_r) =
|R_{PS}(0,T)|^2/|R_{PS}(0,T_r)|^2\cdot K(\tau,M(T),T)/K(\tau,M(T_r),T)$
(note that this expression in general is not normalized to one at $\tau=0$).
However, in Ref.~\cite{Won07} the ratio between the wave functions at the
origin has been neglected: it has a strong temperature dependence and it
is precisely this term that generates the quenching with increasing $T$ of the
ratio in the pseudoscalar channel. Since the mass of the bound states turns out
to be quite stable, the remaining factor, $K(\tau,M(T),T)/K(\tau,M(T_r),T)$, is
nearly constant and yields the flat behavior of the lattice calculations, which
thus here appears just as an artifact of the approximation.

A potential model calculation including the zero modes has also been done in
Ref.~\cite{Moc07}, using a hybrid model where the spectral functions are
calculated by solving a Schr\"odinger equation with a screened potential at low
energies and employing the perturbative expressions at large energies.
In spite of the different models, also this approach yields very good results
for the ratio $G/G{\text{rec}}$ of the $P$-wave correlators, indicating that
this ratio is essentially dominated by the zero mode contribution.
On the other hand, the results of Ref.~\cite{Moc07} in the pseudoscalar channel
are rather different from ours: in the $c\bar{c}$ system at $T$ slightly above
$T_c$, the authors of Ref.~\cite{Moc07} get a rather flat ratio, compatible
with the lattice data; when $T$ grows the lattice ratios get increasingly
quenched, while their calculations show a moderate enhancement
(cf. Fig.~\ref{fig:GratioSNf2} here and Fig.~11 of Ref.~\cite{Moc07}). 
According to the authors of Ref.~\cite{Moc07} the discrepancy should be acribed
to relativistic effects. 

It should be noted that in Ref.~\cite{Moc07} the finite $T$ correlators are
compared to the $T=0$ one and in the latter an ad hoc factor is introduced to
account for radiative corrections: in the pseudoscalar channel this factor
yields a 100\% correction to the low energy part of the spectral function. 
Apparently, no radiative corrections have been included at finite $T$.

To summarize, we have updated a recent calculation \cite{Alb07} of quarkonium
Euclidean correlators at finite $T$, by including the zero mode contributions
that are present in the lattice spectral functions. The model developed in
Refs.~\cite{Alb05,Alb07} is based upon an effective potential extracted from
$Q\bar{Q}$ free and internal energies measured on the lattice.
Good agreement with the lattice results has been found for the $P$-wave
correlators (scalar and axial-vector channels). Also the vector channel appears
to be well described. 

The pseudoscalar correlators, on the other hand, show the same $T$-dependence
as the ones calculated on the lattice, but not the same $\tau$-dependence below
the dissociation temperature. This discrepancy has already been discussed in
Ref.~\cite{Alb07} as being related to the fact that the strength of the bound
states (and of the zero-energy resonances) in the potential model is gradually
decreasing with $T$, whereas the peak associated to the ground state in the
lattice measurements seems to keep a constant strength up to dissociation.
Hopefully, more accurate lattice calculations should shed some ligth on this
issue in the near future.

We would like to stress that these results have been obtained without {\em any}
adjustment of {\em any} parameter, the effective potential being fixed by
independent lattice measurements of the $Q\bar{Q}$ free energies.

\end{document}